\documentclass[12pt,preprint]{aastex}

\newcommand{\be}{\begin{equation}}
\newcommand{\ee}{\end{equation}}
\newcommand{\bea}{\begin{eqnarray}}
\newcommand{\eea}{\end{eqnarray}}

\begin{document}

\title{Grain Acceleration by MHD Turbulence: Gyroresonance Mechanism}

\author{Huirong Yan and A. Lazarian}

\affil{Department of Astronomy, University of Wisconsin, 475 N. Charter
St., Madison, WI 53706; yan, lazarian@astro.wisc.edu}

\begin{abstract}
We discuss a new mechanism of dust acceleration that acts in a turbulent
magnetized medium. The magnetohydrodynamic (MHD) turbulence includes
both fluid motions and magnetic fluctuations. We show that while the
fluid motions bring about grain motions through the drag, the electromagnetic
fluctuations, can accelerate grains through resonant interactions.
In this paper we calculate grain acceleration by gyroresonance in
Cold Neutral Medium. We consider both incompressible and compressible
MHD modes. We show that fast modes dominate grain acceleration. For
the parameters chosen, fast modes render supersonic velocities to
grains that may shatter grains and enable efficient absorption of
heavy elements. Since the grains are preferentially accelerated with
large pitch angles, the supersonic grains get aligned with long axes
perpendicular to the magnetic field. 
\end{abstract}

\keywords{dust, extinction--ISM: particle acceleration--kinematics and dynamics--magnetic
fields}

\section{INTRODUCTION}

Dust is an important constituent that is essential for heating and
cooling of the interstellar medium (ISM). It interferes with observations
in the optical range, but provides an insight to star-formation activity
through far-infrared radiation. It also enables molecular hydrogen
formation and traces the magnetic field via emission and extinction
polarimetry (see Lazarian 2003). The basic
properties of dust (optical, alignment etc.) strongly depend on its
size distribution. The latter evolves as the result of grain collisions,
whose frequency and consequences(coagulation, cratering, shattering,
and vaporization) depend on grain relative velocities (see discussions
in Draine 1985, Lazarian \& Yan 2002a,b).

It is known that all these problems require our understanding of grain
motions in turbulent interstellar medium. Although turbulence has been
invoked by a number of authors (see Kusaka et al. 1970, Draine 1985, Ossenkopf 1993, Weidenschilling \& Ruzmaikina
1994) to provide substantial grain relative motions, the turbulence
they discussed was not magnetized. In a recent paper (Lazarian \&
Yan 2002a, hereafter LY02) we applied the theory of Alfv\'{e}nic
turbulence (Goldreich \& Schridhar 1995, henceforth GS95, Cho, Lazarian
\& Vishniac 2002a for a review) to grain acceleration through gaseous
drag.

Here we account for the acceleration that arises from the resonant
interaction of charged grains with MHD turbulence. To describe the
turbulence statistics we use the analytical fits to the statistics
of Alfv\'{e}nic modes obtained in Cho, Lazarian \& Vishniac (2002b,
hereafter CLV02) and compressible modes obtained in Cho \& Lazarian
(2002, hereafter CL02).

\section{Acceleration of Grains by Gyroresonance}

Turbulent acceleration may be viewed as the acceleration by a spectrum
of MHD waves that can be decomposed into incompressible Alfv\'{e}nic,
and compressible fast and slow modes (see CL02). There exists an important
analogy between dynamics of charged grains and dynamics of comic rays
(see Yan \& Lazarian 2002, henceforth YL02) and we shall modify the
existing machinery used for cosmic rays to describe charged grain
dynamics%
\footnote{In what follows we assume that the time scale for grain charging is
much shorter than the grain Larmor period. %
}. The energy exchange involves resonant interactions between the particles
and the waves. Specifically, the resonance condition is $\omega-k_{\parallel}v\mu=n\Omega$,
($n=0,\pm1,\pm2...$), where $\omega$ is the wave frequency, $k_{\parallel}$
is the parallel component of wave vector ${\textbf{k}}$ along the
magnetic field, $v$ is the particle velocity, $\mu$ is the cosine
of the pitch angle relative to the magnetic field, $\Omega=qB/(mc)$ is the
Larmor frequency of the particle. The sign of $n$ denote the polarization of the wave.
 $+$ represents left hand polarization and $-$ is for right hand polarizarion. Basically there
are two main types of resonant interactions: gyroresonance acceleration
and transit acceleration. Transit acceleration ($n=0$) requires longitudinal
motions and only operates with compressible modes. It happens when
$k_{\parallel}v\mu=\omega$. As the phase speed is $V_{fast}\geq V_{A}$
for fast waves, where $V_{A}$ is the Alfv\'{e}n speed, it is clear
that it can be only applicable to super-Alfv\'{e}nic grains that
we do not deal with here.

Gyroresonance occurs when
the Doppler-shifted frequency of the wave in the grain's guiding center
rest frame $\omega_{gc}=\omega-k_{\parallel}v\mu$ is a multiple of
the particle gyrofrequency, and the rotating direction of wave electric
vector is the same as the direction for Larmor gyration of the grain.
The gyroresonance scatters and accelerates the particles. The efficiency of the two processes for charged grains can be
described by the Fokker-Planck coefficients $D_{\mu\mu}$ and $D_{pp}/p^{2}$,
where $p$ is the particle momentum (see Schlickeiser \& Achatz 1993,
YL02). The ratio of the two rates depends on the ratio of the particle
velocity, the Alfv\'{e}n speed and pitch angle, $p^{2}D_{\mu\mu}/D_{pp}=[(v\zeta/V_{A})+\mu]^{2}$,
where $\zeta=1$ for Alfv\'{e}n waves and $\zeta=k_{\parallel}/k$
for fast modes. We see that the scattering is less efficient for sub-Alfv\'{e}nic
grains unless most particles move parallel to the magnetic field.
We shall show later that as the result of acceleration, $\mu$ will
tend to 0. Therefore in the zeroth order approximation, we ignore
the effect of scattering and assume that the pitch angle cosine $\mu$
does not change while being accelerated. In this case, the Fokker-Planck
equation, which describes the diffusion of grains in the momentum
space, can be simplified (see Pryadko \& Petrosian 1997):

\begin{equation}
\frac{\partial f}{\partial t}+v\mu\frac{\partial f}{\partial z}=\frac{1}{p^{2}}\frac{\partial}{\partial p}p^{2}D_{pp}(\mu)\frac{\partial f}{\partial p},\label{eq:Fokker}\end{equation}
 where $f$ is the distribution function. Apart from acceleration,
a grain is subjected to gaseous friction. For the sake of simplicity
we assume that grains are moving in respect to stationary gas and
turbulence provides nothing but the electromagnetic fluctuations.
The acceleration of grains arising from the gas motion is considered
separately (see LY02). We describe the stochastic acceleration by
the Brownian motion equation: $mdv/dt=-v/S+Y$, where $m$ is the
grain mass, $Y$ is the stochastic acceleration force, $S=t_{drag}/m$
is the mobility coefficient, $t_{drag}$ is the gas drag time (Draine
1985, LY02). The drag time due to collisions with atoms is essentially
the time for collisions with the amount of gas with the mass equal
to that of the grain, $t_{drag}^{0}=(a\rho_{gr}/n_{n})(\pi/8m_{n}k_{B}T)^{1/2}$,
where $a$ is the grain size, $T$ is the temperature, $\rho_{gr}$
is the mass density of grain. We adopt $\rho_{gr}=2.6$gcm$^{-3}$
for silicate grains. The ion-grain cross-section due to long-range
Coulomb force is larger than the atom-grain cross-section. Therefore,
in the presence of collisions with ions, the effective drag time decreases
by the factor given in Draine \& Salpeter (1979). When the grain velocity
gets supersonic (Purcell 1969), the gas drag time is given by $t_{drag}^{s}=t_{drag}/(0.75+0.75c_{s}^{2}/v^{2}-c_{s}^{3}/2v^{3}+c_{s}^{4}/5v^{4})$,
where $c_{s}$ is the sound speed.

Multiply the Brownian equation above by $v$ and take the ensemble
average, we obtain

\begin{equation}
m\frac{d<v^{2}>}{dt}=-\frac{<v^{2}>}{S}+<\dot{\epsilon}>,\label{diff}\end{equation}
 Following the approach similar to that in Melrose (1980), we can
get from Eq.(\ref{eq:Fokker}) the energy gain rate $<\dot{\epsilon}>$
for the grain with pitch angle $\mu$

\begin{equation}
<\dot{\epsilon}>=\frac{1}{p^{2}}\frac{\partial}{\partial p}(vp^{2}D_{pp}(\mu)),\label{epsilon}\end{equation}
 where the Fokker-Planck coefficient $D_{pp}(\mu)$ is given below.

Adopting the result from quasi-linear theory (hereafter QLT, see Schlickeiser
\& Achatz 1993), the momentum diffusion coefficient is (see YL02)

\begin{eqnarray}
D_{pp}(\mu) & = & {\frac{\pi\Omega^{2}(1-\mu^{2})p^{2}V_{A}^{2}}{2v^{2}}}{\mathcal{R}}\int dk^{3}\frac{\tau_{k}^{-1}}{\tau_{k}^{-2}+(\omega-k_{\parallel}v\mu-n\Omega)^{2}}\nonumber \\
 &  & [(J_{n+1}^{2}({\frac{k_{\perp}v_{\perp}}{|\Omega|}})K_{\mathcal{R}\mathcal{R}}({\mathbf{k}}))+J_{n-1}^{2}({\frac{k_{\perp}v_{\perp}}{|\Omega|}})K_{\mathcal{L}\mathcal{L}}({\mathbf{k}}))\nonumber \\
 & - & J_{n+1}({\frac{k_{\perp}v_{\perp}}{|\Omega|}})J_{n-1}({\frac{k_{\perp}v_{\perp}}{|\Omega|}})\nonumber \\
 &  & (e^{i2\phi}K_{\mathcal{R}\mathcal{L}}({\mathbf{k}})+e^{-i2\phi}K_{\mathcal{L}\mathcal{R}}({\mathbf{k}}))],\label{spep}\end{eqnarray}
 where $\tau_{k}$ is the nonlinear decorrelation time and essentially
the cascading time of the turbulence, $k_{\perp}=\sqrt{k_{x}^{2}+k_{y}^{2}}$
is the perpendicular component of the wave vector, $v_{\perp}$ is
the perpendicular component of grain velocity, $\phi=\tan^{-1}(k_{y}/k_{x})$,
$K_{\alpha\beta}(\mathbf{k})$ is the velocity correlation tensor
and will be given in $\S$3, ${\mathcal{L}},{\mathcal{R}}=(x\pm iy)/\sqrt{2}$
represent left and right hand polarization. At the magnetostatic limit, $\tau_{k}\rightarrow\infty$,
the so-called Breit-Wigner-type function transfers into $\delta$
function, i.e., $\tau_{k}^{-1}/(\tau_{k}^{-2}+(\omega-k_{\Vert}v\mu-n\Omega)^{2})\rightarrow\pi\delta(\omega-k_{\Vert}v\mu-n\Omega)$.
Magnetostatic limit is correct for fast moving particles (see YL02),
but for sub-Alfv\'{e}nic grains we should use Eq.(\ref{spep}). However,
we should not integrate over the whole range of $k_{\parallel}$,
because the contribution from large scale is spurious (see YL02).
This contribution stems from the fact that in QLT, an unperturbed
particle orbit is assumed, which results in non-conservation of the
adiabatic invariant $\xi=mv_{\perp}^{2}/2B_{0}$, where $B_{0}$ is
the large-scale magnetic field. Noticing that the adiabatic invariant
is conserved only when the electromagnetic field varies on a time
scale larger than $|\Omega|^{-1}$, we truncate our integral range,
namely, integrate from $k_{res}$ instead of the injection scale $L^{-1}$.
For Alfv\'{e}nic turbulence $\omega=|k_{\parallel}|V_{A}$, the resonant
scale corresponds to $|k_{\parallel,res}|=|\Omega/(V_{A}-v\mu\zeta)|.$
For fast modes in a low $\beta$ medium (where $\beta\equiv P_{gas}/P_{mag}=2c_{s}^{2}/V_{A}^{2}$
is the ratio of gaseous pressure and magnetic pressure), $\omega=kV_{A},$
the resonant scale is $k_{res}=I\Omega/(V_{A}-v\mu\zeta)|$. The upper
limit of the integral $k_{c}$ is set by the dissipation of the MHD
turbulence, which varies with the medium.

Integrating from $k_{res}$ to $k_{c}$, we can obtain from Eq.(\ref{spep})
and (\ref{epsilon}) the energy gain rate $<\dot{\epsilon}>$ as a
function of $v$ and $\mu$. Then with the $<\dot{\epsilon}>$ known,
we can estimate the grain acceleration. Solving the Eq.(\ref{diff})
iteratively, we can get the grain velocity as a function of time.
We check that the grain velocities converge to a constant value after
the drag time. Thus inserting the acceleration rate by fast and Alfv\'{e}n
modes into Eq.(\ref{diff}), we can obtain the final grain velocities
as a function of $\mu$. As $<\dot{\epsilon}>$ increases with pitch
angle, grains gain the maximum velocities perpendicular to the magnetic
field and therefore the averaged $\mu$ decreases.

\section{MHD Turbulence and Its Tensor Description}

Unlike hydrodynamic turbulence,
Alfv\'{e}nic turbulence is anisotropic, with eddies elongated along
the magnetic field. The Alfv\'{e}nic turbulence
is described by GS95 model which postulates that $k_{\bot}v_{k}\sim k_{\parallel}V_{A}$.
This may be viewed as coupling of eddies perpendicular to the
magnetic field and wave-like motions parallel to the magnetic field.
For magnetically dominated, the so-called low $\beta$ plasma, CL02
showed that the coupling of Alfv\'{e}n and compressible modes is
weak and that the Alfv\'{e}n and slow modes follow the GS95 spectrum.
This is consistent with the analysis of HI velocity statistics (Lazarian
\& Pogosyan 2000, Stanimirovic \& Lazarian 2001) .
According to CL02, fast modes are isotropic.
In what follows, we consider both Alfv\'{e}n modes and compressible
modes in low $\beta$ plasma. 

Within the random-phase approximation, the velocity correlation tensor
in Fourier space is (see Schlickeiser \& Achatz 1993) $<v_{\alpha}(\mathbf{k},t)v_{\beta}^{*}(\mathbf{k'},t+\tau)>/V_{A}^{2}=\delta(\mathbf{k}-\mathbf{k'})K_{\alpha\beta}(\mathbf{k})e^{-\tau/\tau_{k}}$,
where $v_{\alpha,\beta}$ is the time-dependent velocity fluctuation
in ${\textbf{k}}$ space associated with the turbulence. 

The velocity correlation tensor for Alfv\'{e}nic turbulence is (CLV02),

\begin{eqnarray}
K_{ij}(\mathbf{k}) & = & \frac{L^{-1/3}}{12\pi}I_{ij}k_{\perp}^{-10/3}\exp(-L^{1/3}|k_{\parallel}|/k_{\perp}^{2/3}),\nonumber \\
\tau_{k} & = & (L/V_{A})(k_{\perp}L)^{-2/3}\sim(k_{\parallel}V_{A})^{-1}\label{anisotropic}\end{eqnarray}
 where $I_{ij}=\{\delta_{ij}-k_{i}k_{j}/k^{2}\}$ is a 2D tensor in
the $x-y$ plane which is perpendicular to the magnetic field, $L$
is the injection scale, $V$ is the velocity at the injection scale.
Velocity fluctuations related to slow modes are subdominant for magnetically
dominated plasmas (CL02) and we do not consider them.

Fast modes are isotropic and have one dimensional energy spectrum
$E(k)\propto k^{-3/2}$ (CL02). In low $\beta$ medium, the velocity
fluctuations are always perpendicular to $\mathbf{B}_{0}$ for all
$\mathbf{k}$, the corresponding correlation is (YL02)\begin{equation}
K_{ij}(\mathbf{k})={\frac{L^{-1/2}}{8\pi}}J_{ij}k^{-7/2},\;\;\tau_{k}=(k/L)^{-1/2}\times V_{A}/V^{2}\label{fast}\end{equation}
 where $J_{ij}=k_{i}k_{j}/k_{\perp}^{2}$ is also a 2D tensor in $x-y$
plane.

\section{Results}

We consider a typical cold neutral medium (CNM), $T=100$K, $n_{n}=30$cm$^{-3},$
$B_{0}=6.3\mu$G. Here, we only consider large grains ($10^{-6}$cm$<a<10^{-4}$cm),
which carry most grains mass ($\sim80\%$) in ISM. The mean grain
charge was obtained from the average electrostatic potential $<U>$
in Weingartner \& Draine (2001). MHD turbulence requires that fluid
velocities are smaller then the Alfv\'{e}n speed. Therefore we assume
that the injection of energy happens at the scale $L$ where the equipartition
between magnetic and kinetic energies, i.e., $V=V_{A}$, is reached.
We assume that the velocity dispersion at the scale $l=10$pc is
5km/s and that the turbulence at large scales
proceeds in tenuous warm media with Alfv\'{e}n speed larger or equal
to 5km/s. In partially ionized medium, a viscosity
caused by neutrals results in decoupling on the characteristic time
scale (see LY02)

\begin{equation}
t_{damp}\sim\nu_{n}^{-1}k^{-2}\sim(l_{n}v_{n})^{-1}k^{-2},\label{cutoff}\end{equation}
 where $\nu_{n}$ is the kinetic viscosity, $l_{n}$ is the neutral
mean free path, $v_{n}$ is the thermal velocity of neutrals. Given
the parameters above, $l_{n}=7\times10^{12}$ cm. When its cascading
rate $\tau_{k}^{-1}=k_{\parallel}V_{A}\sim k_{\perp}^{2/3}L^{-1/3}V_A$ (see Eq.5) equals to the damping rate%
\footnote{As pointed out in $\S$2, the eddy motions happen in the direction
perpendicular to the magnetic field, so $k_{\perp}$is used to calculate
$t_{damp}$ for Alfv\'{e}n modes.%
} $t_{damp}^{-1}(k_{\perp})$, Alfv\'{e}nic turbulence is assumed
damped%
\footnote{Thus we ignore the effect of slowly evolving magnetic structures associated
with a recent reported new regime of turbulence below the viscous
damping cutoff (Cho, Lazarian \& Vishniac 2002c).%
}. This defines the cutoff wave number of the turbulence $k_{\parallel,c}=2.4\times10^{-16}$cm$^{-1}$
and the time scale $\tau_{c}=(k_{\parallel,c}V_{A})^{-1}=1.7\times10^{10}$s.
Assuming that the grain velocities are smaller than Alfv\'{e}n speed,
we can find that the prerequisite for the gyroresonance $|k_{\parallel,c}|>|k_{\parallel\, res}|$
is the same as $\tau_{L}=2\pi/|\Omega|>\tau_{c},$ the condition for
effective hydro drag (see LY02). Thus we see that Alfv\'{e}n modes
cannot accelerate grains (with $a<2\times10^{-5}$cm) through gyroresonance
unless the velocities of these grains are already super-Alfv\'{e}nic.
The cutoff of fast modes corresponds to the scale where the cascading
time scale $\tau_{k}=t_{damp}$ and this gives the cutoff wave number
$k_{c}=4.9\times10^{-15}$cm$^{-1}$. In the present paper we consider
neutral gas of low ionization, and therefore the damping owing to
ions, including collisionless damping, is disregarded (compare to
YL02). Using the procedure described in $\S$2, we obtain the grain
velocities for different pitch angles. The acceleration is maximal
in the direction perpendicular to the magnetic field. Those values
are presented in Fig.1. If averaged over $\mu$, the velocities are
smaller by less than $20\%$.

\begin{figure}
\plotone{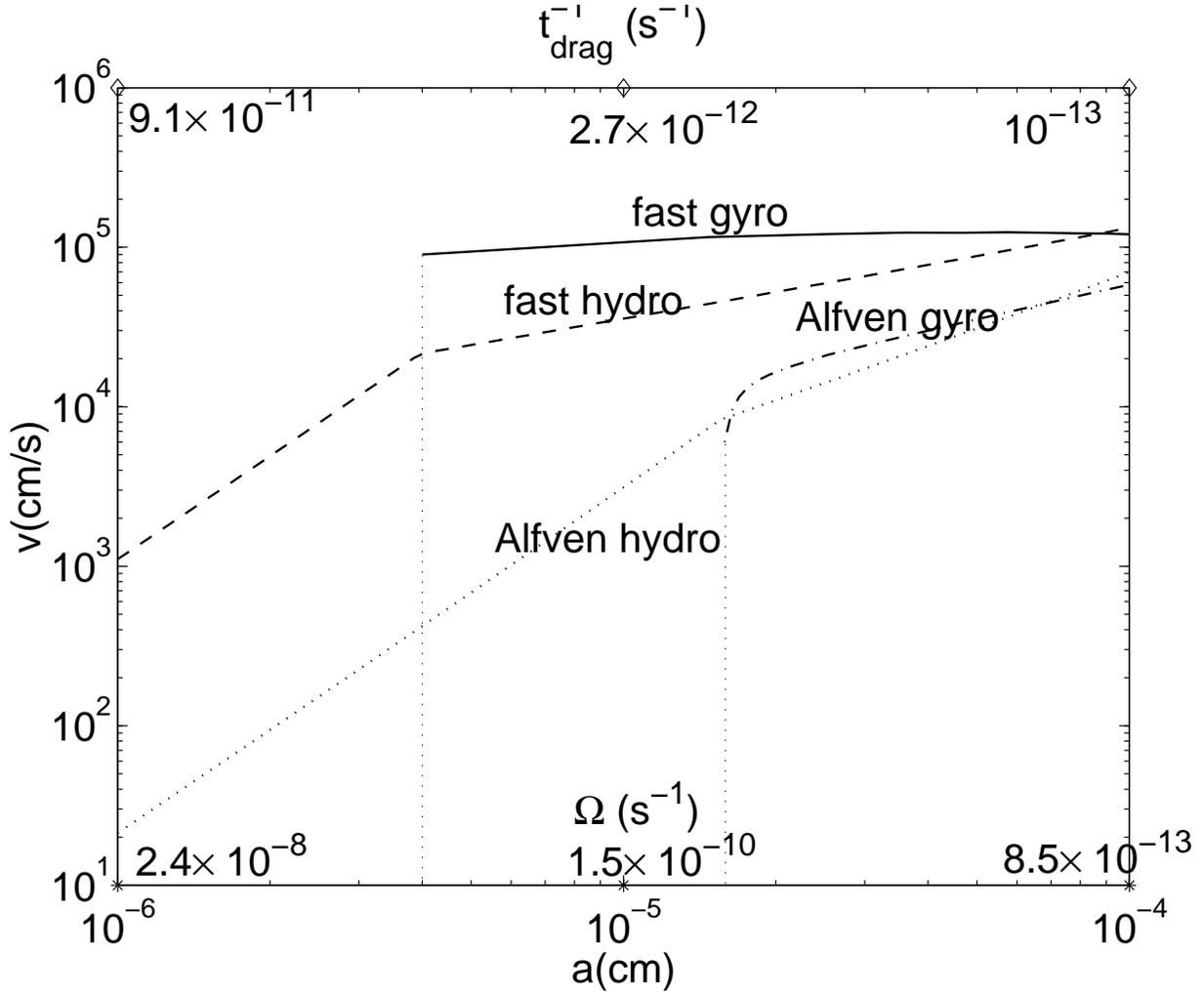}
\caption{Grain velocity vs. size owing to different acceleration processes in CNM.
The solid line represents the gyroresonance with fast modes. The dashdot
line refers to the gyroresonance with Alfv\'{e}n modes. Gyroresonance
with both modes works only for large grains owing to the cutoff by
viscous damping. The cutoff scales for fast and Alfv\'{e}n modes
are different mostly due to the anisotropy of Alfv\'{e}n modes. The
dotted line is the result from hydro drag with Alfv\'{e}n modes (see
LY02), the dashed line represents the hydro drag with fast modes.
The corresponding grain Larmor frequency $\Omega$ and $t_{drag}^{-1}$
are also given along the abscissas.}
\end{figure}

In order to compare different processes, we account for the hydro
drag (see LY02 for a discussion of Alfv\'{e}n-induced drag). The
fast modes cause the relative movement of the grain to the ambient
gas also by gaseous drag. Unlike gyroresonance, this relative motion arises from the decoupling
from the gas. At large scale grains are coupled with the ambient gas,
and the slowing fluctuating gas motions will only cause an overall
advection of the grains with the gas (Draine 1985), which we are not
interested. The largest velocity difference occurs on the largest
scale where grains are still decoupled. While in the hydrodynamic
case the decoupling happens on the time scale $t_{drag}$, the grains
are constrained by the Larmor gyration in MHD case unless $t_{drag}<\tau_{L}$
(see LY02). The latter condition is true for high density gas, however
(see Lazarian \& Yan 2002b). The velocity fluctuations for fast modes
scales as $v_{k}\propto k^{-1/4}\propto\omega^{-1/4}$, where $\omega=kV_{A}$
is the frequency of fast modes (see Eq.(\ref{fast})). In low $\beta$
medium, the velocity fluctuations are perpendicular to the magnetic
field. Therefore, grains get velocity dispersions perpendicular to
the magnetic field, $v\simeq V(\tau_{L}/\tau_{max})^{1/4},$ where
$\tau_{max}=L/V$ is the time scale at the injection scale of turbulence.
Then we also need to consider the effect of damping. Similar to Alfv\'{e}n
modes, the condition for effective hydro drag $\tau_{L}>\tau_{c}=2\pi/\omega_{c}=5\times10^{9}$s
is the same as $k_{c}>k_{res}$ for sub-Alfv\'{e}nic particles, which
is the requirement for gyroresonance. Our calculation shows that the
corresponding grain size is $4\times10^{-6}$cm. For smaller grains,
their velocities are reduced $v\sim v_{c}\times\tau_{L}/\tau_{c}\sim V(\tau_{L}/\tau_{max})^{1/4}(\tau_{L}/\tau_{c})^{3/4}$,
where $v_{c}$ is the velocity of turbulence at the damping scale
(see Lazarian \& Yan 2002b). In Fig.1 we plot the velocity of grain
with $\mu=0$ as a function of grain size since all the mechanisms
preferentially accelerate grains in this direction.

How would the results vary as the parameters of the partially ionized
medium vary? From Eqs.(\ref{diff}), (\ref{epsilon})\&(\ref{spep}), we
find the grain velocity is approximately equal to $v\sim(<\dot{\epsilon}>t_{drag}/m)^{1/2}$$\sim2.5\times10^{4}L_{10}^{-1/4}V_{5}B_{\mu}^{1/2}(q_{e}/a_{5}^{3})^{0.3}t_{drag}^{0.5}$cm/s,
where $L_{10}=L/10$pc is the injection scale defined above, $V_{5}=V_{A}/5{\textrm{km s}^{-1}}$,
$B_{\mu}=B/1\mu$G, $q_{e}=q/1$electron. Noticing that the hydro
drag by fast modes decreases with the magnetic field, $v\propto B^{-1/4}$
beyond the damping cutoff and $v\propto B^{-1}$ below the cutoff
(see the last paragraph), we see that the relative importance of the
gyroresonance and hydro drag depends on the magnitude of the magnetic
field.

It has been shown that the composition of the galactic cosmic ray
seems to be better correlated with volatility of elements (Ellison,
Drury \& Meyer 1997). The more refractory elements are systematically
overabundant relative to the more volatile ones. This suggests that
the material locked in grains must be accelerated more efficiently
than gas-phase ions (Epstein 1980; Ellison, Drury \& Meyer 1997).
The stochastic acceleration of grains, in this case, can act as a
preacceleration mechanism.

Grains moving supersonically can also efficiently vacuum-clean heavy
elements as suggested by observations (Wakker \& Mathis 2000). Grains
can also be aligned if the grains get supersonic (see review by Lazarian
2003). Indeed, the scattering is not efficient for slowly moving grains
so that we may ignore the effect of scattering to the angular distribution
of the grains. Since the acceleration of grains increases with the
pitch angle of the grain (see Eqs.(\ref{epsilon}) and (\ref{spep})),
the supersonic grain motions will result in grain alignment with long
axes perpendicular to the magnetic field.

It is believed that silicate grains won't be shattered unless their
velocities reach $2.7$km/s (Jones et al. 1996). Nevertheless, the
threshold velocities depend on the velocity of turbulence at the injection
scale and on the grain structure, i.e., solid or fluffy. Thus shattering
of the largest grains is possible.

\section{SUMMARY}

In the paper we showed that

1. Fast modes provide the dominant contribution for the acceleration
of charged grains. The velocities obtained are sufficiently high to
be important for shattering large grains and efficiently absorbing
heavy elements from gas. Alfv\'{e}n modes are not important because
of their anisotropy.

2. Depending on the relative importance of the magnetic field, gyroresonance
(strong $B$) or hydro drag (weak $B$) by fast modes dominates grain
acceleration. For small grains, hydro drag by fast modes is the most
important while the gyroresonance is not present as the turbulence
at the resonant frequencies gets viscously damped.

3. In low \(\beta\) medium, all the mechanisms tend to preferentially accelerate grains in
the direction perpendicular to the magnetic field. Among them, gyroresonance
with fast modes can render grains with supersonic motions, which can
result in grain alignment perpendicular to the magnetic field.

\acknowledgements{We acknowledge stimulating communications with E. Zweibel, which
were very important at the initial stage of this work. We thank J.
Mathis, J. Cho and R. Schlickeiser for helpful discussions. This work
is supported by NSF grant AST01-25544.}


\begin{references}

\reference{}Cho, J. \& Lazarian, A. 2002, Phys. Rev. Lett., 88, 245001 (CL02)
\reference{}Cho, J., Lazarian, A. \& Vishniac, E. T. 2002a, in Turbulence and Magnetic field in Astrophysics, Eds. T. Passot \& E. Falgorone (Springer LNP), p.56
\reference{}Cho, J., Lazarian, A. \& Vishniac, E. T. 2002b, ApJ, 564, 291 (CLV02)
\reference{}Cho, J., Lazarian, A. \& Vishniac, E. T. 2002c, ApJ, 566, 49L 
\reference{}Draine, B. T. 1985, in Protostars and Planets II, ed. D. C. Black \&
M. S. Matthews (Tucson: Univ. Arizona Press), p.621 
\reference{}Draine, B. T. \& Salpeter, E. E. 1979, ApJ, 231, 77
\reference{}Ellison, D. C., Drury, L. O'C. \& Meyer, J.-P. 1997, ApJ, 487, 197 
\reference{}Epstein, R. I. 1980, MNRAS, 193, 723
\reference{}Goldreich, P. \& Sridhar, H. 1995, ApJ, 438, 763 (GS95)
\reference{}Jones, A. P., Tielens, A. G. G. M. \& Hollenbach, D. J. 1996, ApJ, 469,
740 
\reference{}Kusaka, T., Nakano, T., \& Hayashi, C., 1970, Prog. Theor. Phys.,
44, 1580 
\reference{}Lazarian, A. 2003, J. Quant. Spectrosc. Radiat. Transfer, 79, 881
\reference{}Lazarian, A. \& Pogosyan, D. 2000, ApJ, 537, 720 
\reference{}Lazarian, A. \& Yan, H. 2002a, ApJ, 566, 105L (LY02)
\reference{}Lazarian, A. \& Yan, H. 2002b, astro-ph/0205283
\reference{}Mathis, J. S. 1990, ARA\&A, 28, 37
\reference{}Melrose, D. B. 1980, Plasma Astrophysics (NY: Gordon \& Breach).
\reference{}Ossenkopf, V. 1993, A\&A 280, 617 
\reference{}Pryadko, J. M. \& Petrosian, V. 1997, ApJ, 482, 774
\reference{}Purcell, E. M. 1969, Physica 41, 100 
\reference{}Schlickeiser, R. \& Achatz, U. 1993, J. Plasma Phy. 49, 63 
\reference{}Stanimirovic, S. \& Lazarian, A. 2001, ApJ, 551, L53
\reference{}Wakker, B. P. \& Mathis, J. S. 2000, ApJ, 544, 107L
\reference{}Weidenschilling, S. J. \& Ruzmaikina, T. V. 1994, ApJ, 430, 713 
\reference{}Weingartner, J. C. \& Draine, B. T. 2001, ApJs, 134, 263 
\reference{}Yan, H. \& Lazarian, A. 2002, Phy. Rev. Lett, 89, 281102 (YL02)
\end{references}
\end{document}